\documentclass[reprint,amsmath,amssymb,aps,pra,twocolumn,longbibliography]{revtex4-2}

\usepackage{upgreek} 
\usepackage{epsfig, amsmath,amsfonts, amssymb,graphicx,amsthm,color}
\usepackage{mathtools}
\usepackage{dcolumn}
\usepackage{bm}
\usepackage{rotating}
\usepackage{units} 
\usepackage{amssymb}
\usepackage{sidecap}
\usepackage{multirow}
\usepackage[american]{babel}
\usepackage{url}
\usepackage[colorlinks=true,        
            linkcolor=blue,         
            citecolor=blue,        
            urlcolor=blue,          
            pdfborder={0 0 0},      
]
            {hyperref}

\begin{document}

\title{Theory of Non-Dichroic Enantio-Sensitive Chiroptical Spectroscopy}
\author{Andr\'es Ord\'o\~nez$^{1}$}
\email[]{andres.ordonez@fu-berlin.de}
\author{David Ayuso$^{2}$, Piero Decleva$^{3}$, Letizia Fede$^{4}$, Debobrata Rajak$^{5}$, Yann Mairesse$^{4}$} 
\author{Bernard Pons$^{4}$}
\email[]{bernard.pons@u-bordeaux.fr}
\affiliation{$^{1}$ Department of Physics, Freie Universit\"at Berlin, 14195 Berlin, Germany}
\affiliation{$^{2}$ Department of Chemistry, Molecular Sciences Research Hub, Imperial College London, SW7 2AZ London, UK}
\affiliation{$^{3}$ Università degli Studi di Trieste, Trieste, Italy}
\affiliation{$^{4}$ Universit\'e de Bordeaux -- CNRS -- CEA, CELIA, UMR5107, Talence, France}
\affiliation{$^{5}$ ELI ALPS, The Extreme Light Infrastructure ERIC, Wolfgang Sandner u. 3., 6728 Szeged, Hungary }

\begin{abstract}
We show that the photoelectron angular distributions produced by elliptical and cross-polarized two-color laser fields interacting with randomly oriented chiral molecules decompose into four irreducible representations of the $D_{2h}$ point group. One of these ($A_u$) corresponds to a non-dichroic enantiosensitive (NoDES) contribution. This NoDES contribution has opposite sign for opposite enantiomers but remains invariant under reversal of the field ellipticity, enabling chirality detection that is robust against variations of the relative phase between orthogonal field components. We propose a protocol to isolate this component using only two velocity-map imaging projections and validate it through numerical simulations. Our calculations, performed in the two-photon resonantly-enhanced ionization, multi-photon, and strong-field ionization regimes with cross-polarized two-color fields show that the NoDES signal reaches about 1\% of the energy-resolved ionization yield, comparable to photoelectron circular dichroism and much larger than standard magnetic-dipole chiroptical effects. NoDES spectroscopy thus provides a symmetry-protected and phase-robust route to probe molecular chirality on the ultrafast time scale. The experimental confirmation of our theory is presented in the companion paper [L. Fede {\em et al.}, arXiv:2512.19062 (2025)].
\end{abstract}

\maketitle
\section{Introduction}
The interaction of chiral molecules with elliptically or circularly polarized light produces a variety of chiroptical effects \cite{berovaComprehensiveChiropticalSpectroscopy2012a}, such as circular dichroism (CD), circularly polarized luminescence \cite{Kumar2015}, magnetochiral circular dichroism \cite{Rikken1997}, and photoelectron circular dichroism (PECD) \cite{ritchie1976,powis2000,sparling2025}, to name a few.
Among them, PECD stands out for two reasons: (i) it reaches high chiral/achiral signal ratios ($\sim10\%$) at the single-molecule level, making it the tool of choice to investigate the time-resolved dynamics of chiral molecules \cite{comby2016,wanie2024} and (ii) it provides direct access to tensorial observables via the multipolar expansion of the photoelectron angular distribution (PAD) $P(\theta,\phi)=\sum_{l,m}b_{l,m}Y_{l,m}(\theta,\phi)$.
The essence of PECD is that photoionization of randomly oriented molecules using circularly polarized light produces a forward-backward asymmetry (FBA), encoded in the odd-$l$ $b_{l,0}$ coefficients.
These odd-$l$ coefficients, which emerge already within the electric-dipole approximation and are rather sensitive to the molecular structure, change sign upon reversal of either the molecular chirality (enantiosensitivity) or the rotation direction of the light polarization (circular dichroism).
The simultaneous enantiosensitive and dichroic behavior is a consequence of symmetry; it follows from how the sample, the polarization of the electric field, and the odd-$l$ spherical harmonics transform with respect to rotations and reflections.
Being rooted in symmetry, PECD is observable across a wide range of molecular species \cite{nahon2015,sparling2025}, ionization regimes \cite{beaulieu2016}, and pulse durations \cite{leePulseLengthDependence2022}.

PECD was initially experimentally investigated using single-photon ionization by XUV radiation \cite{bowering2001,nahon2015}. Later on, it was extended to the multiphoton regime using UV femtosecond laser pulses \cite{lux2012,lehmann2013}, and to the strong-field ionization regime using IR laser pulses \cite{beaulieu2016}.
Accessing the multiphoton and strong-field regimes opened the door to a rich playground in which multiple laser pulses, with various polarization states, can be combined to measure and control the photoionization process \cite{beaulieu2017a,rozen2019,goetzPerfectControlPhotoelectron2019,goetz2019,bloch2021,beaulieu2024,hofmann2024}, as well as to optimize the chiral response \cite{ayuso2022}.
In particular, the nonlinear regime provides access to $b_{l,m}$ coefficients which are absent in the one-photon case due to selection rules.

Selection rules have their origin in symmetries, which can be of geometric or dynamic character.
Geometric selection rules (e.g. $b_{l,m\neq0}=0$ for circularly polarized light due to cylindrical symmetry) are valid independently of the number of photons absorbed in the ionization process.
They simply reflect Curie's principle: the output (PAD) must be as symmetric as the input (sample + polarization).
In contrast, dynamical selection rules (e.g. $b_{l,m}=0$ for $l>2N$, where $N$ is the number of absorbed photons) result from the structure of the integrals describing the specific absorption process.
By moving beyond circular polarization and one-photon absorption, both geometrical and dynamical selection rules can be broken, leading to new $b_{l,m}$ coefficients.
For example, although elliptically polarized light $\vec{E}(t)=E[\cos(\omega t)\hat{x}+\varepsilon\sin(\omega t)\hat{y}]$ breaks cylindrical symmetry, $b_{2,-2}$ ($Y_{2,-2}\propto\sin^{2}\theta\sin2\phi$, we use real-valued spherical harmonics) remains forbidden by a dynamical symmetry in the one-photon case (see e.g. Ref. \cite{ordonezGeometricApproachDecoding2022}).
This dynamical symmetry is broken in the multiphoton case, where $b_{2,-2}$ and other $l$-even $b_{l,m<0}$ coefficients lead to elliptically dichroic 
azimuthal offsets of the PAD in atoms \cite{bashkansky1988,basileTwofoldSymmetricAngular1988,muller1988,lambropoulos1988,nikolopoulosAbovethresholdIonizationNegative1997,manakovEllipticDichroismAngular1999,wangDeterminationCrossSections2000,borcaThresholdEffectsAngular2001,goreslavskiCoulombAsymmetryAboveThreshold2004,popruzhenkoCoulombcorrectedQuantumTrajectories2008,beckerEllipticDichroismStrongfield2022} and molecules \cite{davino2025}.

Over the last decade, the role of elliptically polarized light in photoelectron dichroism has attracted growing attention, leading to the emergence of photoelectron elliptical dichroism (PEELD).
In contrast to the one-photon ionization regime, where the FBA depends linearly on the normalized Stokes parameter $S_{3}/S_{0}=2\varepsilon/(1+\varepsilon^{2})$, in the multiphoton regime this dependence can be non-linear \cite{lux2015,miles2017} and even non-monotonic \cite{comby2018a,sparling2023a,greenwood2023}, providing valuable information about the intermediate resonances in the multiphoton ionization pathway \cite{beauvarlet2022a} and thus enhanced molecular specificity \cite{comby2023}.
Furthermore, besides odd-$l$ $b_{l,0}$ coefficients, recent experiments have reported non-negligible odd-$l$ $b_{l,m>0}$ coefficients that are enantiosensitive and elliptically dichroic \cite{sparling2023a}.

Since $b_{l,m\neq0}$ coefficients describe modulations in the azimuthal angle $\phi$ {[}$Y_{l,m<0}\propto\sin(m\phi)$, $Y_{l,m>0}\propto\cos(m\phi)${]}, they may be lost when taking a single projection of the PAD using a velocity-map imaging spectrometer (VMI) \cite{eppink1997}, as is common in PECD and PEELD studies.
For example, $\cos\phi\propto p_{x}$ is washed out when projected along the $x$ axis.
Indeed, the emergence of $b_{l,m\neq0}$ coefficients rules out the possibility of a full reconstruction of all $b_{l,m}$ coefficients from a single projection \cite{sparling2023a}.
Complete reconstructions can be achieved via direct 3D momentum imaging \cite{rafieefanoodChiralAsymmetryMultiphoton2014,comby2018a,fehre2019b} or through tomographic imaging from multiple projections \cite{wollenhaupt2009,lux2015,comby2018a,sparling2022,sparling2024}.
Nevertheless, as we will show, a pair of projections along judiciously chosen directions are enough to reveal the object of interest in this work, the non-dichroic enantiosensitive (NoDES) signal. 

Enantiosensitive and dichroic $b_{l,m\neq0}$ coefficients have also emerged in photoionization with orthogonal $\omega$-$2\omega$ fields $\vec{E}(t)=E_{\omega}\cos(\omega t)\hat{x}+E_{2\omega}\cos\left(2\omega t+\varphi\right)\hat{y}$ \cite{demekhin2018,demekhinPhotoelectronCircularDichroism2019,rozen2019,rozen2021,bloch2021,ordonez2022}.
These coefficients describe forward-backward up-down asymmetries, i.e. asymmetries such as a $Y_{2,1}\propto p_{x}p_{z}$ quadrupole.
Such asymmetries become evident upon projection of the PAD along the $2\omega$ polarization direction.
An intuitive approach to analyze this type of asymmetry relies on how $\vec{E}(t)$ rotates in opposite directions when $E_{x}(t)>0$ and when $E_{x}(t)<0$ \cite{demekhin2018,rozen2019}. However, the shape and magnitude of the forward-backward up-down asymmetry depend on the phase $\varphi$ between the two frequency components of the field (see Sec. \ref{sec:TDSE-multiphoton-calculations}), as well as on the phases inherent to $\omega$ and 2$\omega$ transition amplitudes \cite{demekhinPhotoelectronCircularDichroism2019}. In spite of this, and from a general point of view, the $C_{2y}$ (rotation by $180^\circ$ around the $y$ axis) symmetry of the field requires any FBA along $z$ to be accompanied by an up-down asymmetry along $x$.
Further symmetry analysis of this field reveals the selection rules governing which $b_{l,m}$ coefficients are allowed, which ones are enantiosensitive, and which ones are dichroic (i.e. change sign upon $\varphi\rightarrow\varphi+\pi$) \cite{ordonez2022}.
Surprisingly, besides $b_{l,m}$ coefficients that are enantiosensitive and dichroic (e.g. $b_{2,1}$), the symmetry analysis predicts a class of $b_{l,m}$ coefficients that are enantiosensitive but non dichroic.
An example is $b_{3,-2}$, which yields an octupolar asymmetry $Y_{3,-2}\propto p_{x}p_{y}p_{z}$, and is predicted to emerge upon absorption of one photon at $\omega$ plus one photon at $2\omega$.
This asymmetry is washed out upon projection of the PAD along the $2\omega$ polarization direction (the most common configuration), but should be visible when projecting along the bisector direction of the two ionizing polarizations \cite{ordonez2022}.
An analogous symmetry analysis reveals that such non-dichroic enantio-sensitive (NoDES) $b_{l,m}$ coefficients also appear for elliptically polarized light.
However, dynamical symmetries rule them out at least in one- and two-photon ionization \cite{ordonez2022}.

Reference \cite{ordonez2022} provided an analytical expression for the lowest-$l$ NoDES coefficient $b_{3,-2}$ in the case of two-photon ionization with orthogonal $\omega$-$2\omega$ fields. However the value of $b_{3,-2}$, or of any other NoDES $b_{l,m}$ coefficient in any ionization regime has not been reported yet.
In a companion letter \cite{companionPRL}, we report on measurement of NoDES signals in the multiphoton and strong-field ionization of chiral molecules by elliptically and orthogonal $\omega$-$2\omega$ laser fields.
In this article, we provide the theoretical framework for NoDES spectroscopy.

In Sec. \ref{sec:Selection-rules-elliptical}, we derive the selection rules for the observation of NoDES $b_{l,m}$ terms in multiphoton ionization by a monochromatic elliptical laser field.
Section \ref{sec:Selection-rules-otc} describes the derivation of these rules for orthogonal two-color laser fields.
Section \ref{sec:Extracting-the-NoDES} shows how the NoDES signal can be extracted using a pair of VMI projections.
In Sec. \ref{sec:Perturbation-theory}, we present simulations of NoDES signals in two-photon ionization of methyloxirane by cross polarized, two-color fields.
Section \ref{sec:TDSE-multiphoton-calculations} describes calculations in the multiphoton and strong field regimes on a toy-model chiral molecule. 
Finally, Sec. \ref{sec:Conclusions} summarizes our conclusions.

\section{Selection rules for elliptical fields}\label{sec:Selection-rules-elliptical}

The PAD $P(\theta,\phi)$ resulting from the interaction between an  elliptically polarized electric field
\begin{equation}
\vec{E}(t)=E_{0}\left[\cos(\omega t)\hat{x}+\varepsilon\sin(\omega t)\hat{y}\right],\label{eq:E_elliptical}
\end{equation}
with a randomly oriented sample of chiral molecules can be split into four contributions:
\begin{equation}
P=P_{NN}+P_{YY}+P_{NY}+P_{YN}\label{eq:PAD_contributions}
\end{equation}
where the first subindex (Y for Yes, N for No) designates enantiosensitivity, and the second subindex dichroism.
That is, $P_{NN}$ is neither enantiosensitive nor dichroic, $P_{YY}$ is both enantiosensitive and dichroic, $P_{NY}$ is not enantiosensitive but is dichroic, and $P_{YN}$ is enantiosensitive but not dichroic.
$P_{YN}$ is the NoDES signal.

Each contribution in Eq. (\ref{eq:PAD_contributions}) is either even or odd with respect to $C_{2x}$ (rotation by $180^{\circ}$ around the $x$ axis) and $\sigma_{xy}$ (reflection in the $xy$ plane) as shown in Table \ref{tab:Characters_elliptical}.
This is because, in the electric-dipole approximation and for randomly oriented
molecules, $C_{2x}$ reverses the ellipticity 
 without changing the enantiomer, while $\sigma_{xy}$ reverses the enantiomer without changing the field.
Furthermore, the PAD must exhibit the joint symmetry of the light-matter system and thus it must be invariant with respect to $C_{2z}$. 

The operators $C_{2z}$, $C_{2x}$ and $\sigma_{xy}$ generate the $D_{2h}$ group.
Each contribution to $P$ in Eq. (\ref{eq:PAD_contributions}) corresponds to a certain irreducible representation of $D_{2h}$, as indicated in parentheses in Table \ref{tab:Characters_elliptical}.
In particular, the NoDES signal corresponds to the $A_{u}$ irreducible representation of $D_{2h}$.
One can extract each of the four contributions $P_{\mathrm{irrep}}=P_{NN},P_{YY},P_{NY},P_{YN}$ by appropriate projection
\begin{equation}
P_{\mathrm{irrep}}=\frac{1}{2^{3}}[1+\chi_{\mathrm{irrep}}^{\sigma_{xy}}\sigma_{xy}][1+\chi_{\mathrm{irrep}}^{C_{2x}}C_{2x}][1+C_{2z}]P \label{eq:projector}
\end{equation}
where $\chi_{\mathrm{irrep}}^{\sigma_{xy}}$ and $\chi_{\mathrm{irrep}}^{C_{2x}}$ are the characters associated to the corresponding contribution (irreducible representation). 

Having identified the relevant irreducible representations of the PAD, it can be expressed in any symmetry-adapted basis to clearly identify the signals according to their enantiosensitivity and dichroism.
For expansions in terms of real spherical harmonics, 
\begin{equation}
P(\theta,\phi)=\sum_{l,m}b_{l,m}Y_{l,m}(\theta,\phi),\label{eq:PAD}
\end{equation}
one can use $C_{2z}Y_{l,m}=(-1)^{m}Y_{l,m}$, $\sigma_{xy}Y_{l,m}=(-1)^{l+m}Y_{l,m}$, $C_{2x}Y_{l,m}=(-1)^{l+m}Y_{l,m}$ if $m\geq0$, and $C_{2x}Y_{l,m}=(-1)^{l+m+1}$ if $m<0$ to assign each $Y_{l,m}$ to the corresponding irreducible representation.
The $b_{l,m}$ coefficients transform accordingly.
This leads to the last two columns of Table \ref{tab:Characters_elliptical}, which provide selection rules for enantiosensitivity and dichroism of the $b_{l,m}$ coefficients.
In particular, the NoDES signal is encoded in the $b_{l,m}$ coefficients with odd $l$ and negative, even $m$.
Table \ref{tab:elliptical-selection-rules-list} lists the enantioselectivity and dichroism of $b_{l,m}$ coefficients with $l\leq9$, which have been experimentally verified in the companion paper \cite{companionPRL}. 

\begin{table}
\begin{centering}
\begin{tabular}{l|rrr|cc}
 & $\sigma_{xy}$ & $C_{2x}$ & $C_{2z}$ & $l$ & $m$\tabularnewline
\hline 
$P_{NN}$ ($A_{g}$) & $1$ & $1$ & $1$ & even & even, $\geq0$\tabularnewline
$P_{YY}$ ($B_{1u}$) & $-1$ & $-1$ & $1$ & odd & even, $\ge0$\tabularnewline
$P_{NY}$ ($B_{1g}$) & $1$ & $-1$ & $1$ & even & even, $<0$\tabularnewline
$P_{YN}$ ($A_{u}$) & $-1$ & $1$ & $1$ & odd & even, $<0$\tabularnewline
\end{tabular}
\end{centering}
\caption{Character table for the different contributions to the PAD {[}Eq.
(\ref{eq:PAD_contributions}){]} resulting from photoionization of randomly oriented chiral molecules using elliptically polarized light {[}Eq. (\ref{eq:E_elliptical}){]}.
The first subindex of $P$ indicates enantiosensitivity and the second dichroism (Y=Yes, N=No).
The corresponding irreducible representations belonging to the group $D_{2h}$ are shown in parenthesis.
The last two columns show the conditions on $l$ and $m$ for real spherical harmonics $Y_{l}^{m}$ belonging to each irreducible representation.
These provide the selection rules for enantioselectivity and dichroism of the corresponding $b_{l,m}$ coefficients {[}Eq. (\ref{eq:PAD}){]}.}
\label{tab:Characters_elliptical}
\end{table}

\begin{table}
\begin{centering}
\includegraphics[width=0.7\linewidth]{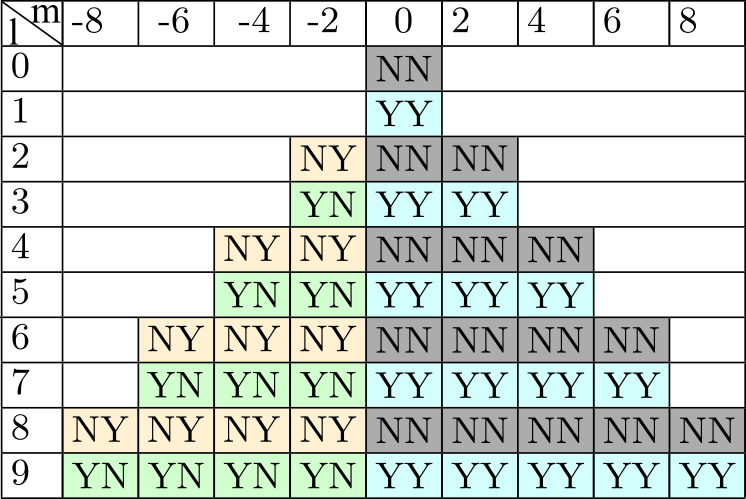} 
\par\end{centering}
\caption{Enantiosensitivity (Y=Yes, N=No) and dichroism (Y/N) of the $b_{l,m}$ coefficients {[}see Eq. (\ref{eq:PAD}) and Table \ref{tab:Characters_elliptical}{]} obtained upon ionization of randomly oriented chiral molecules with elliptically polarized light for $l\protect\leq9$ {[}Eq. (\ref{eq:E_elliptical}){]}.
See experimental confirmation in the companion paper \cite{companionPRL}.
}\label{tab:elliptical-selection-rules-list}
\end{table}

For circular polarization, $|\varepsilon|=1$, and we must replace $C_{2,z}$ by $C_{\infty,z}=\{C_{\phi}|\phi=[0,2\pi)\}$, the set of all rotations around $z$. Thus we deal with $D_{\infty h}$ instead of $D_{2h}$. According to the correlation between the irreducible representations of $D_{\infty h}$ and $D_{2h}$ ($\Sigma_{g}^{+}\leftrightarrow A_{g}$, $\Sigma_{u}^{+}\leftrightarrow B_{1u}$, $\Sigma_{g}^{-}\leftrightarrow B_{1g}$, $\Sigma_{u}^{-}\leftrightarrow A_{u}$), for circular polarization $P_{NN}=\Sigma_{g}^{+}$, $P_{YY}=\Sigma_{u}^{+}$, $P_{NY}=\Sigma_{g}^{-}$, $P_{YN}=\Sigma_{u}^{-}$. It is easy to see that $\Sigma_{g}^{-}$ and $\Sigma_{u}^{-}$ cannot be represented using single-particle functions, like spherical harmonics, and thus there is neither a $P_{NY}$ nor a $P_{YN}$ contribution in this case. Indeed, since cylindrical symmetry only allows $m=0$, the effect of $\sigma_{xy}$ and $C_{2x}$ on the allowed spherical harmonics is identical, i.e. $\sigma_{xy}Y_{l,0}=C_{2x}Y_{l,0}=(-1)^{l}Y_{l,0}$. Thus enantiosensitivity and dichroism are represented by the same (odd-$l$) harmonics. This “degeneracy” is lifted in the less symmetric $D_{2h}$ group, which allows for $m\neq 0$, and different transformation rules for $Y_{l,m}$ under $\sigma_{xy}$ and $C_{2x}$.

Interestingly, the $P_{NY}$ and $P_{YN}$ contributions could emerge for circular polarization when additional degrees of freedom are resolved (e.g. electron and ion, or electron momentum and spin, in coincidence). In this case the PAD can be expanded in terms of two-particle functions, namely coupled spherical harmonics $\mathcal{Y}_{L,M=0}^{l_{1}l_{2}}(\theta_{1},\phi_{1};\theta_{2},\phi_{2})$. Their transformation rules with respect to $\sigma_{xy}$ and $C_{2x}$ allow for $P_{NY}$ and $P_{YN}$ (NoDES) contributions -- see Ref. \cite{Suzuki2024} for a related discussion in the context of recoil-frame photoelectron angular distributions in one-photon ionization with unpolarized light. We remark that detecting two-particle angular distributions \cite{Fehre2021} is considerably more demanding than the single-particle case discussed here and that the presence of NoDES contributions in that context has not been verified experimentally yet.

\section{Selection rules for orthogonal two-color fields}\label{sec:Selection-rules-otc}

\begin{table}
\begin{centering}
\begin{tabular}{l|rrr|cc}
 & $\sigma_{xy}$ & $C_{2x}$ & $C_{2y}$ & $l$ & $m$\tabularnewline
\hline 
$P_{NN}$ ($A_{g}$) & $1$ & $1$ & $1$ & even & even, $\geq0$\tabularnewline
$P_{YY}$ ($B_{2g}$) & $-1$ & $-1$ & $1$ & even & odd, $\ge0$\tabularnewline
$P_{NY}$ ($B_{2u}$) & $1$ & $-1$ & $1$ & odd & odd, $<0$\tabularnewline
$P_{YN}$ ($A_{u}$) & $-1$ & $1$ & $1$ & odd & even, $<0$\tabularnewline
\end{tabular}
\end{centering}
\caption{Same as Table \ref{tab:Characters_elliptical} but for photoionization with an orthogonal two-color field {[}Eq. (\ref{eq:otc}){]}.
}
\label{tab:Characters_OTC}
\end{table}

\begin{table*}
\begin{centering}
\includegraphics[width=0.7\linewidth]{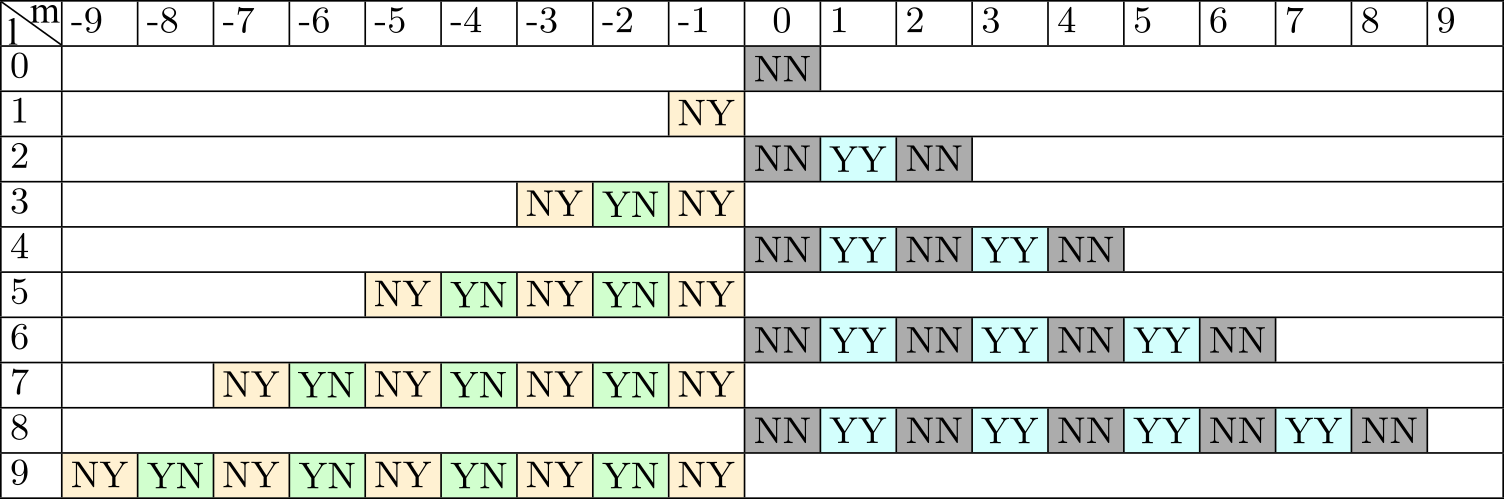}
\par\end{centering}
\caption{
Enantiosensitivity (Y=Yes, N=No) and dichroism (Y/N) of the $b_{l,m}$ coefficients {[}see Eq. (\ref{eq:PAD}) and Table \ref{tab:Characters_OTC}{]} obtained upon ionization of randomly oriented chiral molecules with an orthogonal two-color field {[}Eq. (\ref{eq:otc}){]} for $l\protect\leq9$.
See numerical confirmation in Sec. \ref{sec:TDSE-multiphoton-calculations} and experimental evidence in the companion paper \cite{companionPRL}. 
}\label{tab:TableOTC}
\end{table*}

In the case of an orthogonal two-color field
\begin{equation}
\vec{E}(t)=E_{0}\left[\cos(\omega t)\hat{x}+r\cos(2\omega t+\varphi)\hat{y}\right],\label{eq:otc}
\end{equation}
we can also express the PAD as in Eq. (\ref{eq:PAD_contributions}).
In this case, the light-matter system is invariant with respect to $C_{2y}$ instead of $C_{2z}$, and dichroism is understood as a change of sign upon $\varphi\rightarrow\varphi+\pi$.
$P_{NN}$ and $P_{YN}$ remain the same as in the elliptical case, while $P_{YY}$ and $P_{NY}$ now correspond to the $B_{2g}$ and $B_{2u}$ irreducible representations, respectively, as shown in Table \ref{tab:Characters_OTC}.
The conditions on $l$ and $m$ are obtained using $C_{2y}Y_{l,m}=(-1)^{l}Y_{l,m}$ if $m\geq0$ and $C_{2y}Y_{l,m}=(-1)^{l+1}Y_{l,m}$ if $m<0$.
Note that since the NoDES signal is again encoded in the $A_{u}$ irreducible representation, the corresponding $b_{l,m}$ coefficients are the same as in the elliptical case.
The projector on a given irreducible representation $P_{\mathrm{irrep}}$ can be obtained by replacing $C_{2z}\rightarrow C_{2y}$ in Eq. (\ref{eq:projector}).
Table \ref{tab:TableOTC} lists the enantioselectivity and dichroism of $b_{l,m}$ coefficients with $l\leq9$, which are numerically verified in Sec. \ref{sec:TDSE-multiphoton-calculations}.
Experimental evidence of the presence of the NoDES signal ($P_{YN}$) for this field, obtained using the approach presented in the next section, is presented in the companion paper \cite{companionPRL}.

\section{Extracting the NoDES signal using two VMI projections }\label{sec:Extracting-the-NoDES}

As shown in previous sections, the NoDES signal is encoded in the $A_{u}$ irreducible representation of $D_{2h}$, which corresponds to $b_{l,m}$ coefficients with odd $l$ and even negative $m$.
These are associated to real spherical harmonics $Y_{l,2\mu}\propto\sin(2\mu\phi)$, $\mu=1,2,\dots,(l-1)/2$.
Due to the $\sin(2\mu\phi)$ dependence, such contributions vanish upon integration over $p_{x}$ or $p_{y}$, and thus the NoDES signal is washed out in standard VMI projections.
While the NoDES signal can be recovered by doing full 3D reconstructions of the PAD, such reconstructions can be quite demanding.
In Ref. \cite{ordonez2022} it was suggested to project the PAD along the $\hat{x}+\hat{y}$ direction, to avoid washing out the NoDES signal. However, such projection mixes the NoDES contributions ($P_{YN}$) with the other three ($P_{NN}$, $P_{YY}$, $P_{NY}$).
Here we show that by taking the difference between the $\hat{x}+\hat{y}$ projection and the $\hat{x}-\hat{y}$ projection, one can reveal a pure NoDES signal.
In practice, instead of projecting on two different directions, one rotates the field polarization with respect to the VMI detector, which leads to a corresponding rotation of the PAD.
Thus, our approach is the following:
For the first (second) projection, the field is rotated by $\pi/4$ ($-\pi/4$) radians around the $z$ axis, and the projection of the rotated PAD is done along $y$.
This yields $P_{\pm\pi/4}(p_{x},p_{z})$.
The difference of the two projections is taken as $\Delta P(p_{x},p_{z})\equiv [P_{\pi/4}(p_{x},p_{z})-P_{-\pi/4}(p_{x},p_{z})]/2$.
The 2D NoDES signal $P_{\mathrm{NoDES}}(p_{x},p_{z})$ can be extracted from the forward-backward asymmetric part of $\Delta P(p_{x},p_{z})$, i.e. $P_{\mathrm{NoDES}}(p_{x},p_{z}) \equiv [\Delta P(p_{x},p_{z}) - \Delta P(p_{x},-p_{z})]/2$.
To see this, consider the rotated distributions projected along $p_{y}$
\begin{align}
P_{\pm\pi/4}(p_{x},p_{z}) & \equiv\int_{-\infty}^{\infty}\mathrm{d}p_{y}C_{8z}^{\pm}P(p_{x},p_{y},p_{z})\nonumber \\
 & =\sum_{l,m}\int_{-\infty}^{\infty}\mathrm{d}p_{y}b_{l,m}(p)Y_{l,m}(\theta,\phi\mp\frac{\pi}{4}),
\end{align}
where $C_{8z}^{\pm}$ is the operator that rotates functions by $\pm\pi/4$ around the $z$ axis.
The spherical harmonics can be written as $Y_{l,m}(\theta,\phi)=\mathcal{P}_{l,\left|m\right|}(\cos\theta)G_{m}(\phi)$, where $\mathcal{P}_{l,\left|m\right|}(\cos\theta)$ are (suitably normalized) generalized Legendre functions and $G_{0}(\phi)=1$, $G_{|m|}(\phi)=\cos(|m|\phi)$ and $G_{-|m|}=\sin(|m|\phi)$. Therefore, the difference between the rotated PADs is given by 
\begin{align}
  \Delta P(p_{x},p_{z}) = \sum_{l,m}\int_{-\infty}^{\infty}\mathrm{d}p_{y}b_{l,m}(p)\mathcal{P}_{l,\left|m\right|}(\cos\theta)\Delta G_{m}(\phi),
\end{align}
where $\Delta G_{m}(\phi)\equiv [G_{m}(\phi-\pi/4)-G_{m}(\phi+\pi/4)]/2$.
Since $p=\sqrt{p_{x}^{2}+p_{y}^{2}+p_{z}^{2}}$ and $\cos\theta=p_{z}/p$ are even functions of $p_{y}$, the parity of the integrand is determined by $\Delta G_{m}(\phi)$, which is given by 
\begin{align}
\Delta G_{m}(\phi) =\begin{cases}
\sin\left(\frac{m\pi}{4}\right)\sin(m\phi) & m>0\\
0, & m=0\\
\sin\left(\frac{m\pi}{4}\right)\cos(m\phi) & m<0
\end{cases}
\end{align}
Polar plots of $\sin(m\phi)$ and $\cos(m\phi)$ show that they are odd and even in $p_{y}$, respectively. Thus, integrals of $m>0$ terms in the expression for $\Delta P(p_{x}, p_{z})$ vanish, and only $m<0$ terms contribute to $\Delta P(p_{x},p_{z})$ (except multiples of $4$).
Looking at Tables \ref{tab:Characters_elliptical} and \ref{tab:Characters_OTC}, we see that $\Delta P (p_{x},p_{z})$ only contains contributions from $P_{NY}$ and $P_{YN}$, which contain $m<0$ terms.
Furthermore, since $P_{NY}$ and $P_{YN}$ contributions are respectively even and odd in $p_{z}$, according to their characters with respect to $\sigma_{xy}$ in Tables \ref{tab:Characters_elliptical} and \ref{tab:Characters_OTC}, it follows that 
\begin{align}
P_{\mathrm{NoDES}}\left(p_{x},p_{z}\right) &\equiv\frac{1}{2} \left[\Delta P(p_{x},p_{z})-\Delta P(p_{x},-p_{z})\right]\nonumber \\
 & =\int_{-\infty}^{\infty}\mathrm{d}p_{y}\Delta P_{YN}(p_{x},p_{y},p_{z})
\end{align}
where $\Delta P_{YN}\equiv (C_{8z}^{+}P_{YN}-C_{8z}^{-}P_{YN})/2$ depends exclusively on the $P_{YN}$ (NoDES) contribution.
We demonstrate this approach for multiphoton and strong-field ionizations with orthogonal two-color fields in the numerical simulations in Sec. \ref{sec:TDSE-multiphoton-calculations}, and experimentally in the companion paper \cite{companionPRL}.

\section{Perturbation theory: resonantly enhanced two-photon ionization of methyloxirane}\label{sec:Perturbation-theory}

The selection rules derived in Sections \ref{sec:Selection-rules-elliptical} and \ref{sec:Selection-rules-otc} allow the emergence of NoDES terms, but do not provide information on their magnitude.
In this section, we consider the scheme initially proposed in \cite{ordonez2022}, i.e. two-photon ionization using a linearly polarized ultraviolet pulse and its orthogonally polarized second harmonic.
We consider methyloxirane, a molecule where PECD has been extensively studied \cite{stenerDensityFunctionalStudy2004,strangesValencePhotoionizationDynamics2005,tommasoConformationalEffectsCircular2006,continiVibrationalStateDependence2007,garciaVibrationallyInducedInversion2013,rafieefanoodChiralAsymmetryMultiphoton2014,garciaIdentifyingUnderstandingStrong2017a}.
The energy difference between ground and first excited electronic states is $\Delta E=7.12$ eV \cite{cohenExcitedElectronicStates1983}, while the ionization energy from the ground state is $I_{p}=10.4$ eV \cite{strangesValencePhotoionizationDynamics2005}.
Thus, two-photon resonantly enhanced ionization $|0\rangle\xrightarrow{2\omega}|1\rangle\xrightarrow{\omega}|k\rangle$ using $2\omega=7.12$ eV and $\omega=3.56$ eV ($\lambda=348$ nm) leads to photoelectrons with kinetic energy $E_{k}=-I_{p}+2\omega+\omega=0.28$ eV.
We model the excitation using the highest occupied and lowest unoccupied molecular orbitals (HOMO and LUMO) to represent the states $|0\rangle$ and $|1\rangle$, respectively. We neglect the contribution from all other non-resonant intermediate states, assume a fixed-nuclei approximation, and use second-order perturbation theory.
Under these approximations, the value of $b_{3,-2}/b_{0,0}$ depends only on the transition dipoles and is given by \cite{ordonez2022,ordonezGeometricApproachDecoding2022}
\begin{widetext}
\begin{equation}
\frac{b_{3,-2}}{b_{0,0}}=\frac{1}{4}\sqrt{\frac{15}{7}}\frac{\int\mathrm{d}\Omega_{k}\left[\hat{k}\cdot\left(\vec{d}_{1,0}\times\vec{d}_{\vec{k},1}^{*}\right)\right]\left[5\left(\hat{k}\cdot\vec{d}_{\vec{k},1}\right)\left(\hat{k}\cdot\vec{d}_{1,0}\right)-\left(\vec{d}_{\vec{k},1}\cdot\vec{d}_{1,0}\right)\right]}{\int\mathrm{d}\Omega_{k}\left(2\left|\vec{d}_{\vec{k},1}\right|^{2}\left|\vec{d}_{1,0}\right|^{2}-\left|\vec{d}_{\vec{k},1}\cdot\vec{d}_{1,0}\right|^{2}\right)},\label{eq:b3m2_b00}
\end{equation}
\end{widetext}
where $\vec{d}_{1,0}\equiv\langle1|\vec{d}|0\rangle$ and $\vec{d}_{\vec{k},1}\equiv\langle\vec{k}|\vec{d}|1\rangle$ are the relevant transition dipoles, $|\vec{k}\rangle$ is the scattering state describing a photoelectron with momentum $\vec{p}=\hbar\vec{k}$ in the molecular frame, $\hat{k}$ is the unit vector in the direction of $\vec{k}$, and $\int\mathrm{d}\Omega_{k}$ is an integral over all photoelectron directions.
This expression was derived using the approach presented in Ref. \cite{ordonezGeometricApproachDecoding2022}. Note that in the framework of lowest-order perturbation theory used here, $b_{3,-2}$ does not depend on $\varphi$ \cite{ordonez2022}.

\begin{figure}
\begin{centering}
\includegraphics[width=0.9\linewidth]{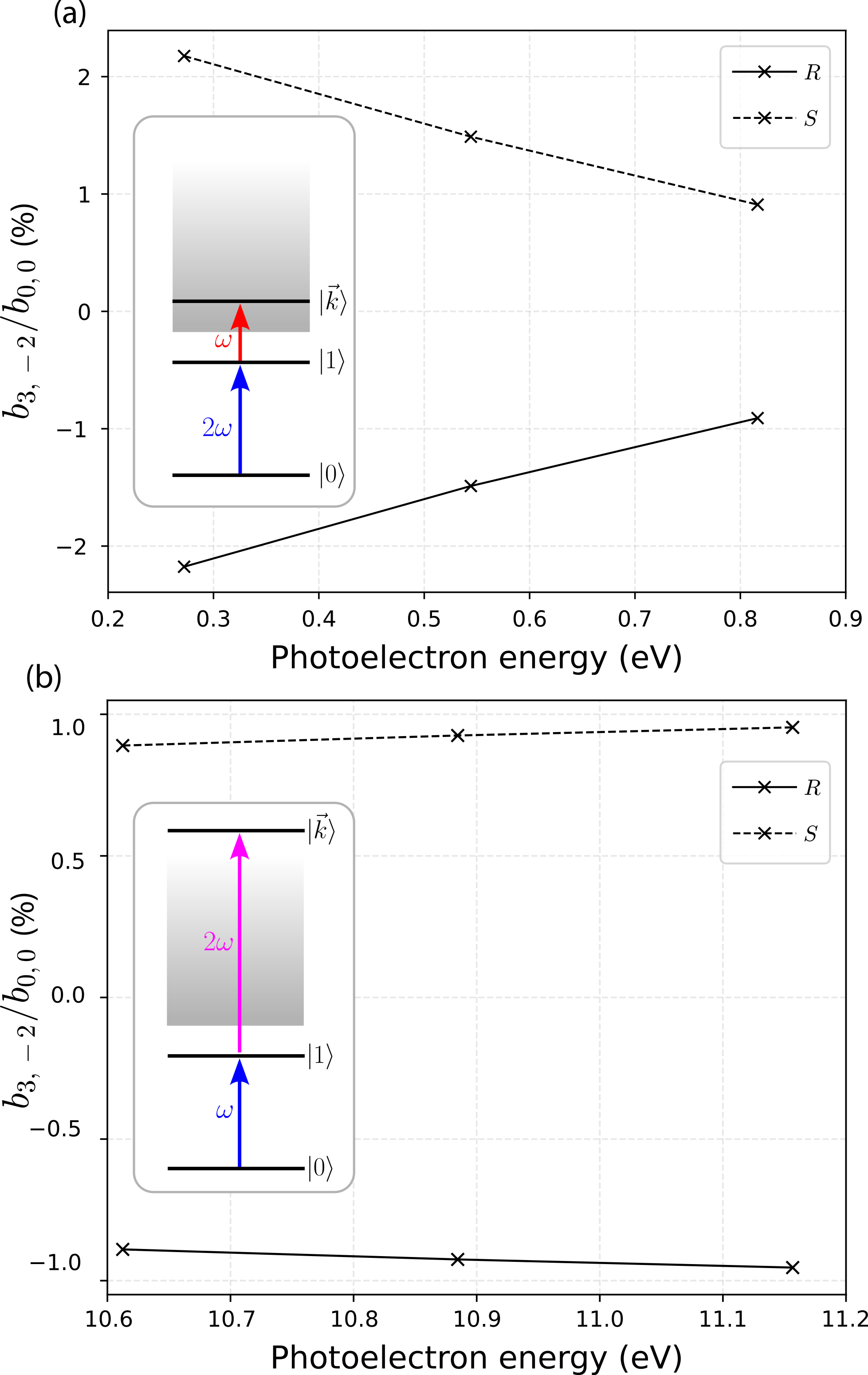} 
\par\end{centering}
\caption{Perturbation-theory calculations for the lowest NoDES contribution $b_{3,-2}$ normalized to the total yield $b_{0,0}$ {[}Eq. (\ref{eq:b3m2_b00}){]} as a function of the photoelectron energy in two-photon resonantly enhanced ionization of methyloxirane with an orthogonal two-color field {[}Eq. (\ref{eq:otc}){]} for two different photon orderings.
In (a) $\omega=3.56$ eV while in (b) $\omega=7.12$ eV.
The ground and intermediate states are described by the HOMO and LUMO orbitals.
}\label{fig:b3m2_perturbative}
\end{figure}

We calculated the transition dipoles using state-of-the-art methods \cite{toffoliConvergenceMulticenterBspline2002,stranges2005,toffoliTiresiaCodeMolecular2023}, which often lead to quantitative agreement with PECD experiments.
We also consider the process $|0\rangle\xrightarrow{\omega}|1\rangle\xrightarrow{2\omega}|k\rangle$ using $\omega=7.12$ eV ($\lambda=174$ nm), which leads to photoelectrons with kinetic energy $E_{k}=-I_{p}+2\omega+\omega=10.96$ eV.
Note that changing the photon ordering also induces an extra minus sign in Eq. (\ref{eq:b3m2_b00}), see Appendix E of Ref. \cite{ordonezGeometricApproachDecoding2022}.
Figure \ref{fig:b3m2_perturbative} shows $b_{3,-2}/b_{0,0}$ for a few values of the photoelectron kinetic energy in low and high photoelectron energy regions.
$b_{3,-2}/b_{0,0}$ is of the order of $1$\% in both regions, which is similar to the ratio $b_{1,0}/b_{0,0}$ reached in multiphoton PECD of methyloxirane \cite{rafieefanoodChiralAsymmetryMultiphoton2014}, and is considerably higher than traditional CD chiroptical signals ($\sim$0.01\% \cite{tanakaCircularlyPolarizedLuminescence2018}).
Since the value of $b_{3,-2}/b_{0,0}$ is highly dependent on the transition dipole matrix elements, exploring different intermediate states and final photoelectron energies can potentially enhance it \cite{goetz2019}. We have not attempted such optimization.

\section{TDSE calculations: multiphoton and strong-field ionizations of a toy-model chiral molecule}\label{sec:TDSE-multiphoton-calculations}

In this section, we analyze the emergence of NoDES structures in the multiphoton and strong-field ionization regimes where the perturbative approach becomes impractical. We thus model the interaction by direct integration of the Time-Dependent Schrödinger Equation (TDSE) for a toy-model chiral molecule with ionization potential $I_{p}=8.98$ eV, as detailed in \cite{rozen2019}.
Despite the simplicity of the molecular structure (made of four nuclei and one electron), this model was shown to produce signals qualitatively resembling the results of experiments performed in fenchone, in various field configurations (orthogonal two-color bilinear fields \cite{rozen2019,rozen2021}, co-rotating \cite{bloch2021} and counter-rotating \cite{beaulieu2024} two-color bicircular fields).
Here we let the molecule interact with a fundamental 800 nm laser field of intensity $I$ and an orthogonally polarized 400 nm laser field, with an amplitude ratio $r=30\%$ and a phase shift $\varphi$ between the two components of the electric field.
The laser pulse duration is four cycles of the fundamental, with one-cycle ascending and descending ramps.
The short duration induces carrier-envelope dependent effects, whose influence is canceled by appropriate symmetrization of the photoelectron distributions, as detailed in \cite{rozen2019}.
The calculation is averaged over random molecular orientations to obtain the final photoelectron momentum distribution. Practical details on the numerical recipe can be found in \cite{rozen2019}.
As in PECD, the chiral potential induces an FBA in the electron scattering process, and the enantiosensitive part of the PAD is thus obtained by extracting the FBA: $A(p_{x},p_{y},p_{z})=\frac{1}{2}\left[P(p_{x},p_{y},p_{z})-P(p_{x},p_{y},-p_{z})\right]$.

\begin{figure}
\centering{}\includegraphics[width=1\linewidth]{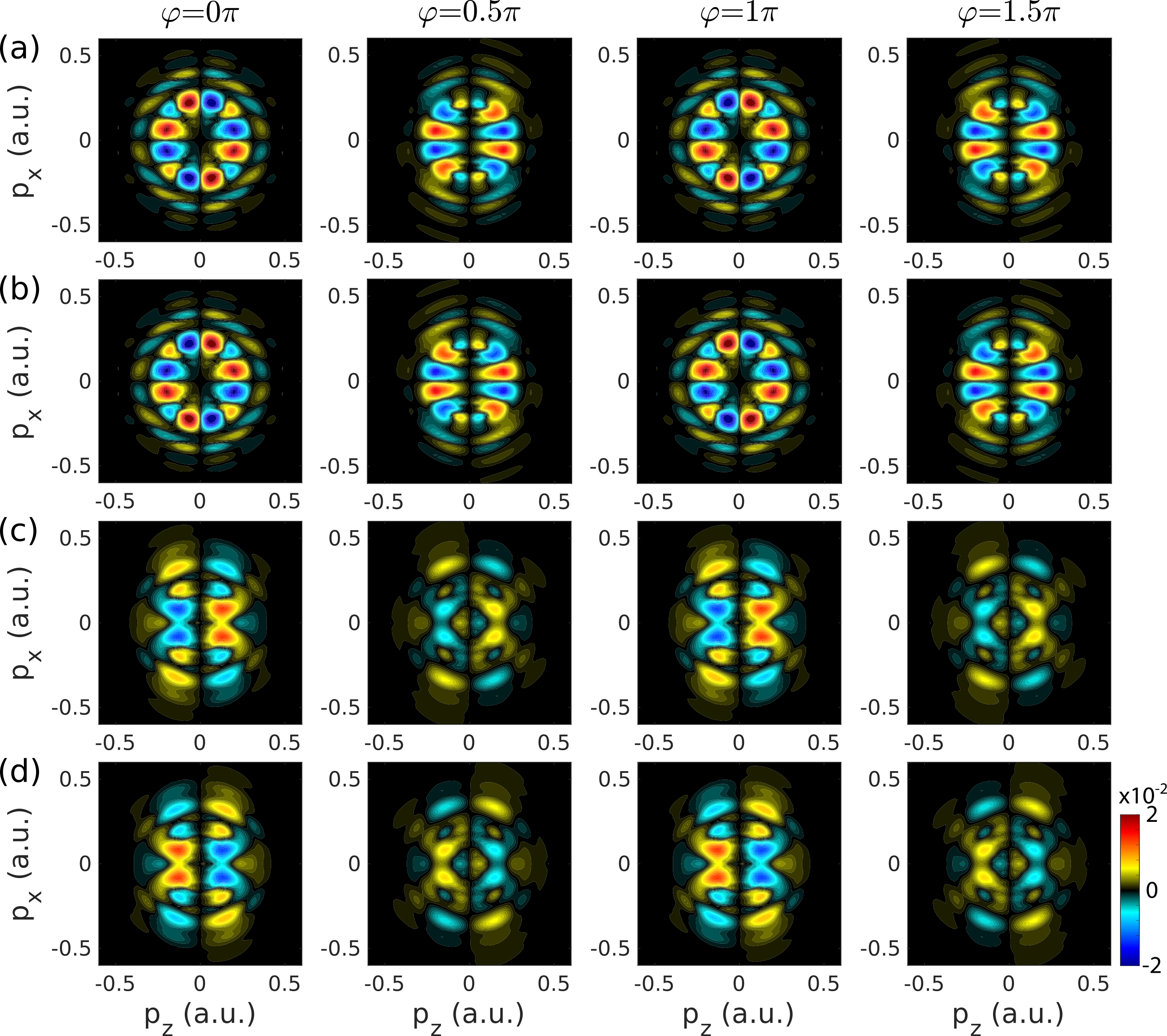} \caption{TDSE calculations of the photoionization of a toy-model chiral molecule by an orthogonally polarized two-color laser field at various relative phases $\varphi$ [Eq. (\ref{eq:otc})].
The intensity of the 800 nm fundamental field component is $I=5\times 10^{12}$ W/cm$^2$ so that the interaction lies in the multiphoton regime. 
(a-b) Projections of the forward-backward antisymmetric part of the PAD along the $\hat{y}$ direction, in one enantiomer (a) and its mirror image (b).
(c-d) NoDES signal obtained by subtracting the $\hat{x}+\hat{y}$ and $\hat{x}-\hat{y}$ projections of the forward-backward antisymmetric part of the PAD (see Sec. \ref{sec:Extracting-the-NoDES}), in one enantiomer (c) and its mirror image (d).  }\label{FigTDSE}
\end{figure}

We first consider the case of a fundamental 800 nm field with $I=5\times 10^{12}$ W/cm$^2$. The Keldysh parameter associated to the interaction is $\gamma=3.9$, which is characteristic of the multiphoton ionization regime.  
Figure \ref{FigTDSE} (a,b) shows projections of the FBA in the ($x,z$) plane as a function of $\varphi$, obtained in the two enantiomers of the chiral molecule.
The FBA is up-down antisymmetric, reflecting the $C_{2y}$ symmetry of the field and the switch of the effective laser ellipticity every half laser period, as predicted and observed in previous studies \cite{demekhin2018,rozen2019}, as well as in the measurements reported in \cite{companionPRL}.
The FBA changes sign when switching enantiomer, and also when the phase between the two colors is shifted by $\pi$, i.e. when the ellipticity of the ionizing field reverses.
The observed signal is thus enantiosensitive and dichroic.
Indeed, the non-dichroic part of the enantiosensitive signal is expected to cancel out in the projection along the $\hat{y}$ direction.
To reveal it, we follow the method presented in Sec. \ref{sec:Extracting-the-NoDES}, and used in the experiment in the companion paper \cite{companionPRL}.
We project the FBA along the $\hat{x}+\hat{y}$ and $\hat{x}-\hat{y}$ directions, and plot half the difference between the two projections.
The results (Fig. \ref{FigTDSE}(c,d)) show a significant FBA, in the $1\%$ range of the maximum electron yield, and keep a constant sign upon shifting $\varphi$ by $\pi$.
This term is thus non-dichroic.
Comparing the results obtained in opposite enantiomers shows that it changes sign, confirming that this procedure enables detecting the NoDES component. 

\begin{figure*}
\centering{}\includegraphics[width=0.75\linewidth]{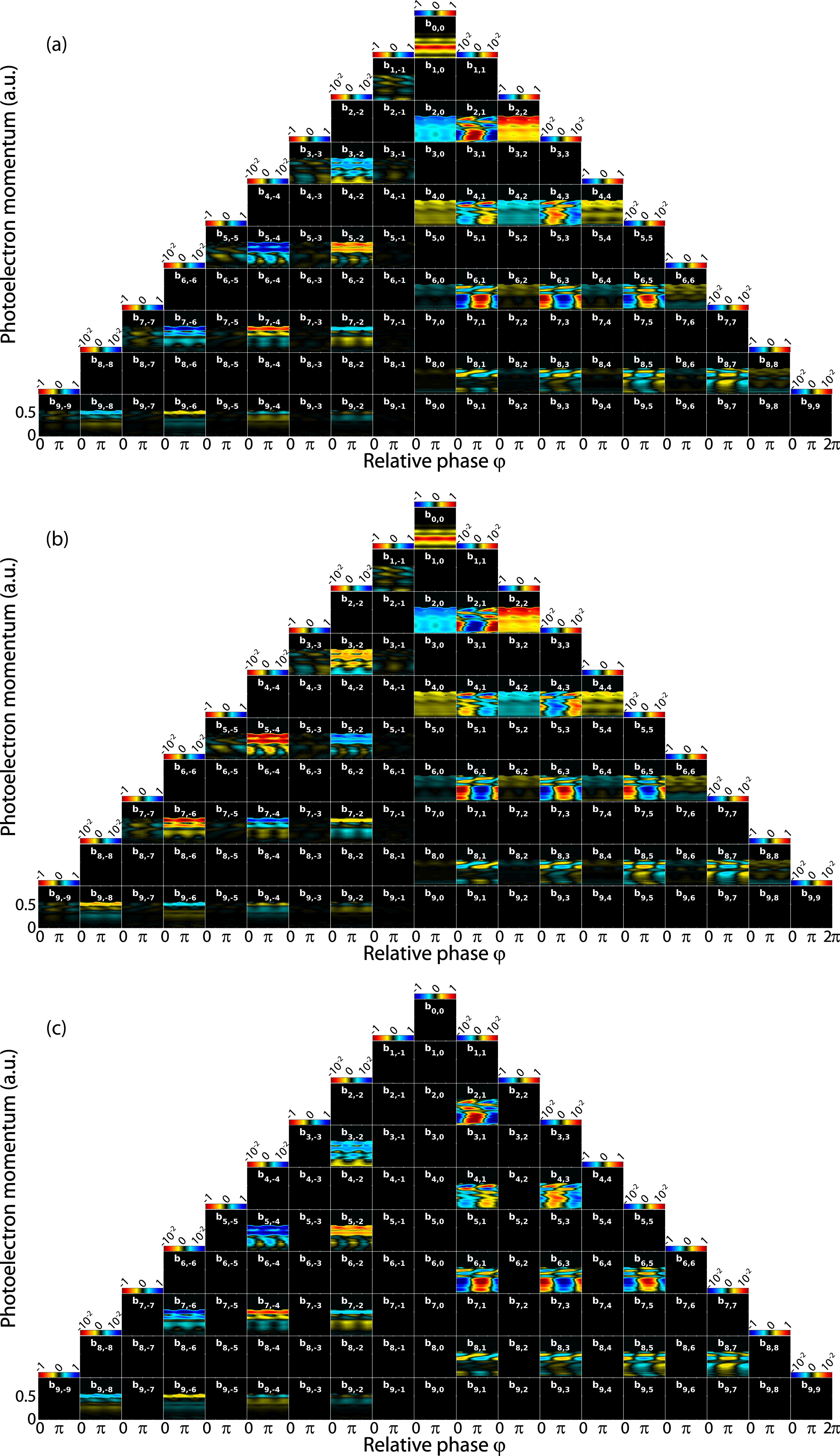} \caption{Spherical harmonic decomposition of the photoelectron angular distributions obtained by TDSE simulations of the photoionization of a toy-model chiral molecule (first enantiomer in (a), second enantiomer in (b)) by an orthogonal two-color field at $5\times10^{12}$ $\mathrm{W/cm}^{2}$.
The enantiodifferential, forward-backward antisymmetric signal is shown in (c).}
\label{FigTDSESH}
\end{figure*}

The spherical harmonic decomposition of the 3D PAD should give direct access to the NoDES terms.
We thus apply the fitting procedure \cite{politis2016} used for experimental strong-field PEELD distributions in \cite{companionPRL}, to the simulated distributions.
Figure \ref{FigTDSESH} shows the coefficients of the different spherical harmonics as a function of $\varphi$ in both enantiomers as well as the enantiodifferential coefficients.
The decomposition was performed using $l_{max}=9$.
The $l\neq0$ coefficients are normalized by $b_{0,0}$.
The distribution shows two classes of terms, with different orders of magnitude: the even $m\geq0$ terms and the odd $m<0$ ones are dominant, while the odd $m>0$ terms and the even $m<0$ ones are two orders of magnitude weaker.
Switching enantiomer (Fig. \ref{FigTDSESH} (a-b)) leaves the former terms unchanged and changes the sign of the latter.
The enantiosensitive terms are isolated in the enantiodifferential signals (Fig. \ref{FigTDSESH} (c)).
The odd $m>0$ components oscillate around zero as a function of $\varphi$, changing sign when $\varphi$ is shifted by $\pi$.
These terms are thus dichroic.
By contrast almost all the $m<0$ terms show a constant sign as a function of $\varphi$.
Some components, like the low-momentum part of $b_{5,-2}$, change sign as a function of $\varphi$, but with a double periodicity: they switch sign when $\varphi$ is shifted by $\pi/2$, and thus have the same sign when $\varphi$ changes by $\pi$.
These terms are thus non-dichroic.
The sign changes they may show is related to the change in the shape of the laser field, which evolves from a ``C'' shape at $\varphi=0$ to a ``8'' shape at $\varphi=\pi/2$.
Thus, the enantiosensitivity and dichroic nature of the different spherical harmonic coefficients we found in these simulations confirms the symmetry analysis leading to Table \ref{tab:TableOTC}. 

\begin{figure}
\centering{}\includegraphics[width=1\linewidth]{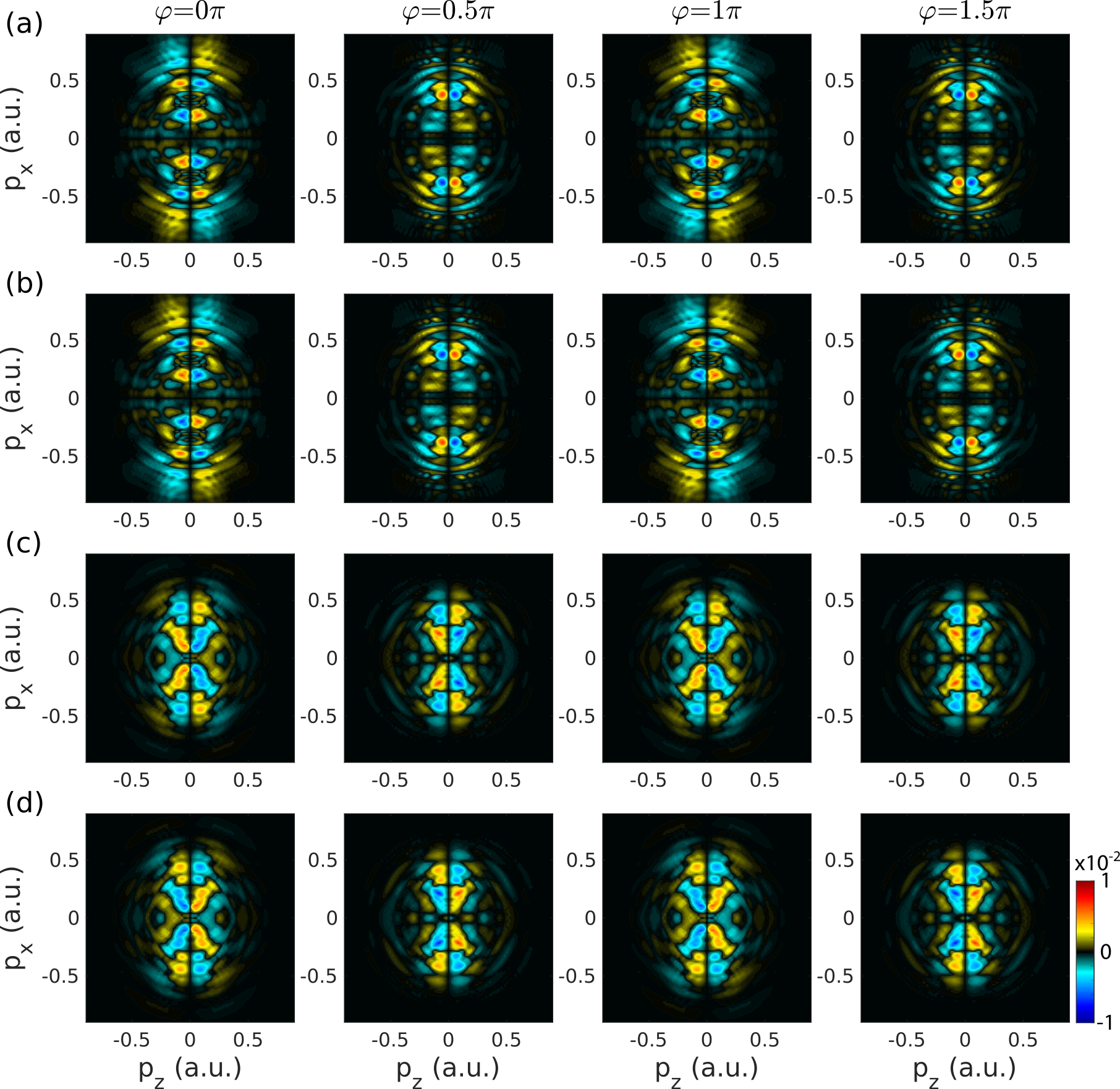} \caption{Same as Fig. 2 but for $I=5\times 10^{13}$ W/cm$^2$ so that the interaction lies in the strong-field regime.} \label{FigTDSE_StrongField}
\end{figure}

Next, we consider an 800 nm laser field with intensity $I=5\times 10^{13}$ W/cm$^2$ to enter the strong-field ionization regime. The Keldysh parameter associated to the interaction is now $\gamma=1.2$, which is characteristic of (non-adiabatic) tunneling ionization processes. Fig. \ref{FigTDSE_StrongField}(a,b) presents the projections of the FBA along the $\hat{y}$ direction as a function of $\varphi$, for the two enantiomers of the toy-model molecule. The FBA is stretched along the $x$ direction compared to the multiphoton case of Fig. \ref{FigTDSE} and exhibits higher above-threshold ionization peaks. This is consistent with strong-field driven electron dynamics. The FBA is up-down antisymmetric, following the sub-cycle ellipticity of the two-color laser field and its $C_{2y}$ symmetry, and it changes sign when the phase $\varphi$ is increased by $\pi$. It also changes sign when the enantiomeric form of the molecular sample is reversed. Therefore, the whole FBA, projected along $y$, is dichroic and enantiosensitive, as in the multiphoton case. We implemented again the procedure presented in Sec. \ref{sec:Extracting-the-NoDES} to isolate the NoDES signal. The results are presented in Fig. \ref{FigTDSE_StrongField}(c,d) for both enantiomers and varying $\varphi$. The non-dichroic nature of the signal shows up through the up-down symmetry of all the results and their insensitivity to $\varphi\rightarrow\varphi+\pi$. The enantiosensitivity corresponds to the antisymmetry of the graphs displayed in Fig. \ref{FigTDSE_StrongField}(c) and Fig. \ref{FigTDSE_StrongField}(d). Beyond its insensitivity to $\varphi \rightarrow \varphi + \pi$ phase variation, the NoDES signal does not depend as strongly as the rest of the FBA on smaller variations of $\varphi$, as evidenced in Fig. \ref{FigTDSE_StrongField} through the comparison of FBA and NoDES results at $\varphi=0$ and $\varphi=\pi/2$. 

\begin{figure*}
\centering{}\includegraphics[width=0.75\linewidth]{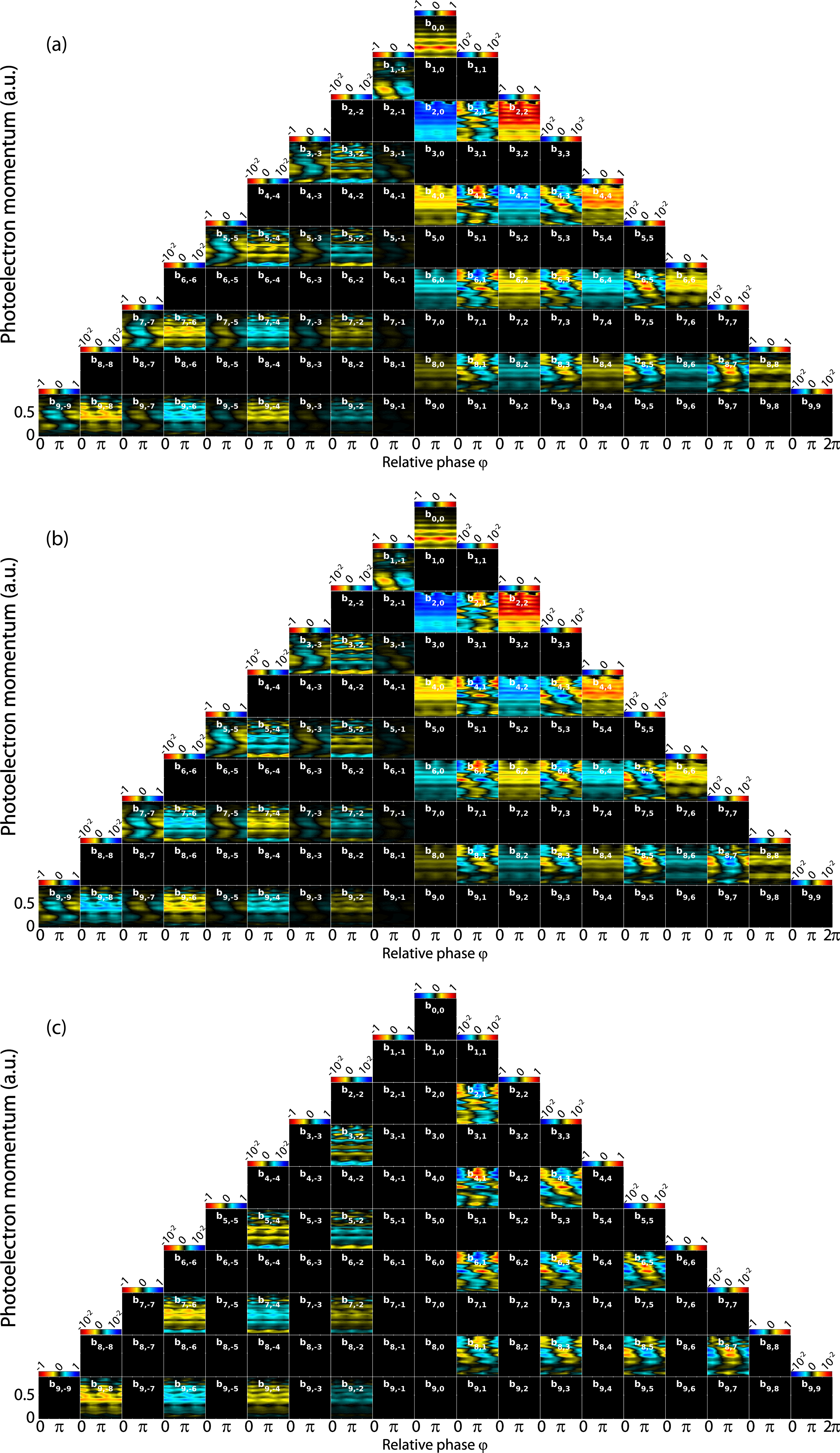} \caption{
Same as Fig. 3 but for $I=5\times 10^{13}$ W/cm$^2$.} \label{FigTDSESH_StrongField} 
\end{figure*}

Finally, we undertook spherical harmonic decompositions of the 3D PADs associated to the strong-field interaction. The results are presented in Fig. \ref{FigTDSESH_StrongField} for the toy-model molecule (a), its mirror image (b) and their differences in (c). The results are again fully consistent with Table IV derived from symmetry arguments: the PAD consist of even $l$, $m \geq 0$ and odd $l$, $m<0$ components. Within the former group, the non-dichroic and non-enantiosensitive coefficients, $b_{l,m}$ with even $l$ and $m$, have a larger relative contribution to the PAD in the strong-field regime than in the case of multiphoton ionization (see e.g. $b_{6,m\geq 0}$ in Fig. \ref{FigTDSESH} (a) and Fig. \ref{FigTDSESH_StrongField}(a)). Within the odd $l$, $m<0$ group, the dichroic and non-enantiosensitive components, $b_{l,m}$ with odd $l$ and $m$, also exhibit a larger magnitude with respect to the NoDES ones in the strong-field configuration (see e.g. $b_{7,m<0}$ in Fig. \ref{FigTDSESH} (a) and Fig. \ref{FigTDSESH_StrongField}(a)). This may be due to the fact that the electric field progressively inhibits the role of the ionic potential in the electron dynamics as the field intensity increases. However, the NoDES contributions, $b_{l,m}$ with odd $l$ and even $m$, still present significant ($\sim 1\%$) magnitude, up to rather high photoelectron kinetic energies and large angular momenta.

\section{Conclusions}\label{sec:Conclusions}

We have shown how the symmetry properties of elliptical and orthogonal two-color fields interacting with randomly oriented chiral molecules leads to photoelectron angular distributions that can be decomposed into four irreducible representations of the $D_{2h}$ point group.
One of these representations, $A_u$, describes a component of the photoelectron angular distribution that changes sign when the enantiomer is replaced by its mirror image, but not when the ellipticity of the field is reversed.
This non-dichroic enantiosensitive (NoDES) component enables the detection of molecular chirality in a way that is robust against variations in the relative phase between orthogonal field components.
Since this robustness stems from the symmetry properties of the system, NoDES signals are an example of symmetry protection, as observed in other fields of physics \cite{plotnikExperimentalObservationOptical2011, guTensorentanglementfilteringRenormalizationApproach2009}, in the context of chiroptical spectroscopy. 

We have provided a measurement protocol to extract the NoDES component of the photoelectron angular distribution, without background from the other components, using only two velocity-map imaging projections.
This protocol has been tested both on numerical simulations here, and on experimental data in the companion paper \cite{companionPRL}. 

We have quantified the magnitude of the NoDES signal for orthogonal two-color fields in perturbative two-photon, multiphoton and strong-field ionization regimes, and found that it is of the order of 1\% of the energy-resolved ionization yield.
This order of magnitude is in agreement with the experimental data in the companion paper \cite{companionPRL}.
Furthermore, this magnitude is comparable to that of photoelectron circular dichroism in the strong-field regime, and thus much bigger than standard chiroptical effects relying on magnetic-dipole interactions, which are typically below $0.1\%$ \cite{tanakaCircularlyPolarizedLuminescence2018}.
Therefore, NoDES spectroscopy can be used to image molecular chirality on the ultrafast electronic time scale in a broad range of ionization regimes, and in the absence of a fixed relative phase between the orthogonal field components.

\begin{acknowledgments}
  We acknowledge Val\'erie Blanchet and Baptiste Fabre for fruitful discussions.
  A. O. acknowledges funding from the Deutsche Forschungsgemeinschaft (DFG, German Research Foundation) - 543760364.
  D.A. acknowledges funding from the Royal Society URF$\backslash$R$\backslash$251036.
  This work is part of the ULTRAFAST and TORNADO projects of PEPR LUMA and was supported by the French National Research Agency, under grant under ANR-21-CE30-038-01 (Shotime), and as a part of the France 2030 program, under grants ANR-23-EXLU-0002 and ANR-23-EXLU-0004.
  We acknowledge the financial support of the IdEx University of Bordeaux / Grand Research Program ``GPR LIGHT''.
  This project has received funding from European Union's Horizon 2020 research and innovation programme under grant agreement no. 871124 Laserlab-Europe.
  The ELI ALPS project (GINOP-2.3.6-15-2015-00001) is supported by the European Union and co-financed by the European Regional Development Fund. 
\end{acknowledgments}
\nocite{apsrev4-2Control}

\bibliographystyle{apsrev4-2}
\bibliography{biblioPRA}

\end{document}